\documentclass[reprint,superscriptaddress,amsmath,amssymb,aps]{revtex4-2}
\usepackage{subfig}
\usepackage{graphicx}
\usepackage{dcolumn}
\usepackage{bm}
\usepackage{amsmath}
\usepackage{tabularx}
\usepackage{epsfig}
\usepackage{rotating}
\usepackage[font=small,labelfont=bf,justification=raggedright,format=plain]{caption}
\usepackage[version=4]{mhchem}
\usepackage{enumerate}
\usepackage{xcolor}
\begin{document}

\preprint{APS/123-QED}

\title{Understanding coupled mass-heat transport in fluids by approach-to-equilibrium molecular dynamics}
\author{Antonio Cappai}
  \affiliation{Department of Physics, University of Cagliari, Cittadella Universitaria, I-09042 Monserrato (CA), Italy}
\author{Luciano Colombo}
  \affiliation{Department of Physics, University of Cagliari, Cittadella Universitaria, I-09042 Monserrato (CA), Italy}
 \author{Claudio Melis}
 \affiliation{Department of Physics, University of Cagliari, Cittadella Universitaria, I-09042 Monserrato (CA), Italy}

\date{\today}

\begin{abstract}
We present a generalization of AEMD approach, routinely applied to estimate thermal conductivity, to the more general case in which Soret and Dufour effects determine a coupled heat-mass transfer. We show that, by starting from microscopical definitions of heat and mass currents, conservation laws dictates the form of the differential equations governing the time evolution. In particular, we focus to the well specific case in which a closed-form solution of the system is possible and derive the analytical form of time-evolution of temperature and concentration scalar fields in the case in which step-like initial profiles are imposed across a rectangular simulation cell. The validity of this new generalized expression is finally validated using as benchamrk system a two-component Lennard-Jones liquid system, for which generalized diffusivities are estimated in different reduced temperature and density region of phase diagram.
\end{abstract}

\maketitle

\section{\label{sec:1}Introduction}
Thermal transport in fluids is a subject of great interest, both for fundamentals physics and applications \cite{PhysRevLett.123.138001,gunawan2013liquid}. In particular, whenever thermal and mass fluxes occur simultaneously a coupled transport regime is established, as reported since the nineteen century (with the process of thermal diffusion, known as the Ludwig-Soret effect, first observed in 1856 \cite{ludwig1856difusion,soret1879etat}) and rationalized by Non-Equilibrium Thermodynamics \cite{de2013non,kjelstrup2008non}.\par
Several computational techniques have been devised to the aim of quantifying the magnitude of thermal and mass coupling transport coefficients, hereafter referred to as Soret and Dufour coefficients, based on Equilibrium or Non-Equilibrium approaches and formulated in terms of Onsager coefficients. In the first case, by properly taking into consideration the microscopically observed correlation between mass and heat fluxes, Green-Kubo based approaches have been successfully used to estimate the coupling parameters \cite{he2012lattice}. Alternatively, non-equilibrium simulations have been used by imposing a thermal gradient and next analyzing the steady state mass concentration profile \cite{bonella2017thermal}. One of the major bottleneck of these techniques, however, is related to the fact that very long simulation times are indeed needed in order to obtain reliable data, an issue especially challenging if the size of the system is large \cite{sellan2010size,schelling2002comparison,muller1997simple}.\par

In this study, we present a novel approach not requiring long simulations based on a generalization of the well-established Approach to Equilibrium methodology extensively used for thermal conductivity calculations in solid systems. We focus on the simultaneous diffusion of mass and thermal energy in a model binary mixture, and we provide a physically sound formal device, which we prove to be theoretically robust, easy to implement, and computationally efficient to estimate coupling transport coefficient.\par
We proceed under the less restrictive conditions to establish a formal description of the relaxation towards equilibrium of an initial non-equilibrium profile of temperature and mass density. By generating the relaxation dynamics through a computer simulation, we get information on the coefficients governing the strength of heat/mass coupling. Our model binary mixture is described by the Lennard-Jones (LJ) potential, which is general enough to cover a plethora of possible real liquid mixtures.

The paper is structured as follows. In Section II the phenomenological picture of the transport phenomena (Fourier regime, Fick regime, Soret/Dufour coupled regimes) is sketched, and the constitutive hypothesis of our model, with no assumptions except for the liquid and binary nature of the mixture, are presented. In Section III, the very general hypothesis previously stated are used in order to describe the relaxation of an imposed step-like temperature and density gradient in a Lennard-Jones binary liquid. The conditions and approximations introduced to formally treat the coupled regime are kept at minimum, with the only aim to allow the derivation of an analytical form for the solutions. By introducing the concept of generalized diffusivity a decoupling of the partial differential equations is eventually achieved. In this context, the solution for the transient evolution of the spatially averaged temperature difference and concentration differences is worked out. In Section IV the computational framework allowing the actual generation of temperature and concentration profiles is presented: the numerical implementation is performed using an equimolar mixture of LJ fluids as benchmark system.\par
Here we as well present and discuss the results of our simulations, namely the estimate of thermal diffusivities, diffusion coefficients and the coupling parameters: as byproduct of the assessment of the method, the demonstration of the absence of a rectification effect in a Lennard Jones liquid mixture is eventually achieved.

\section{\label{sec:3}Phenomenological picture}
\begin{figure*}
    \centering
    \includegraphics[scale=0.2]{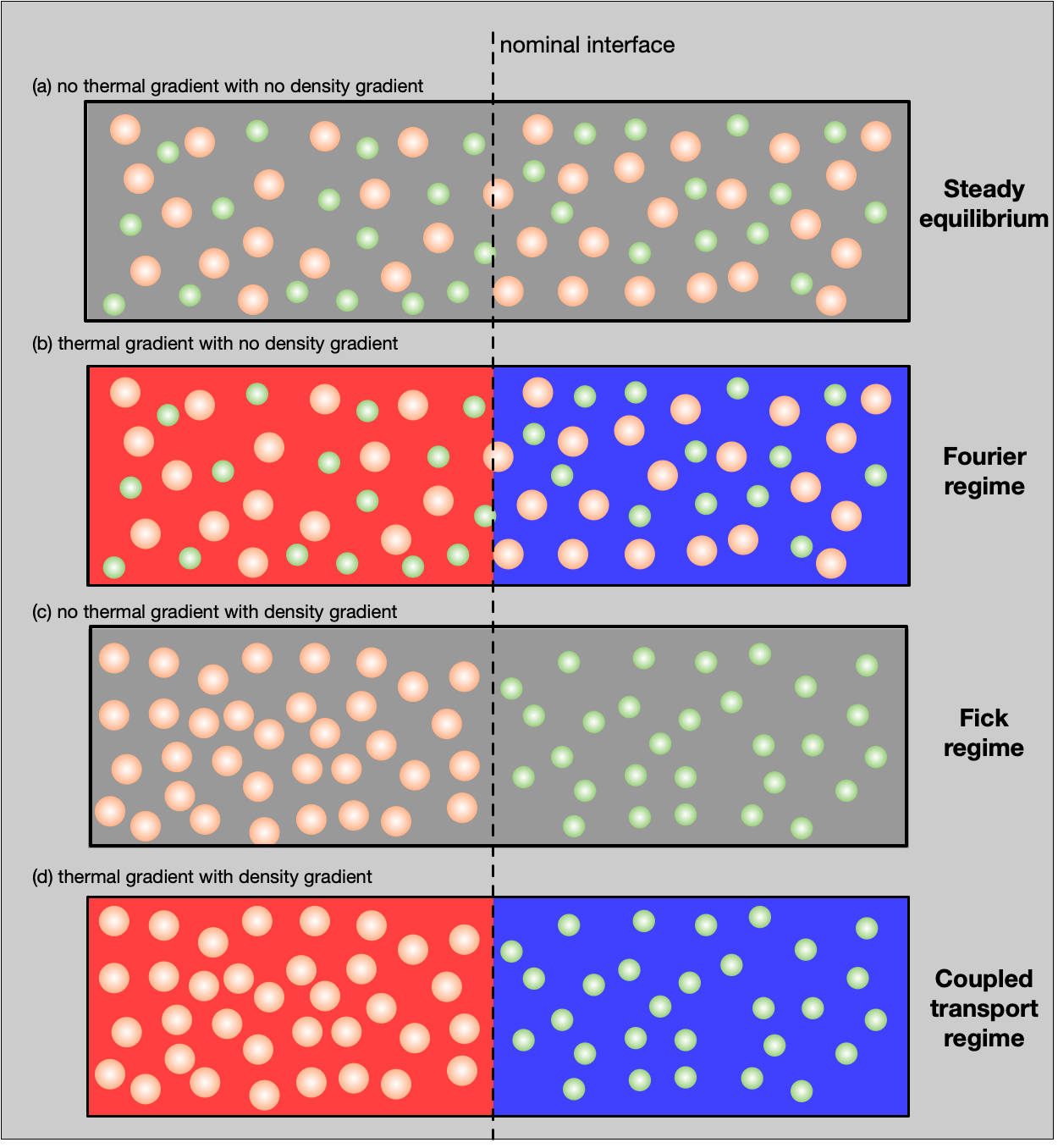}
    \caption{The four possible cases: no temperature and no mass density gradient; temperature gradient and no mass density gradient, pure Fourier regime; no temperature gradient, mass density gradient, pure Fick regime, temperature and mass density gradient simultaneously present, coupled regime}
    \label{fig:enter-label}
\end{figure*}
The phenomena we address in this paper are namely the \emph{Soret} and \emph{Dufour effects}. We understand that we will consider fluid systems, i.e. systems that can carry a mass current.\par
It is common knowledge that if a temperature gradient is present in a homogeneous and isotropic sample, a corresponding heat flux is established and the relation linking the temperature gradient $\vec { \nabla} T$ to the corresponding heat current $\vec { J}_q$ is linear
\begin{equation}
\vec{ J}_q=\frac{dQ}{dA dt}\hat{ n}=\kappa \vec { \nabla} T
\end{equation}
where $dQ$ represents the amount of heat flowing from high to low temperature regions per unit cross section $dA$ in unit of time $dt$ along the direction marked by the unit vector $\hat{ n}$. When this equation is obeyed we will address the situation as \emph{Fourier regime} where the coefficient $\kappa$ is the thermal conductivity of the system.\par
A similar relation exists in the case of a non-homogeneous system, at constant temperature, if the mass density $\rho$ is non uniformly distributed: in this case, a mass current $\vec{ J}_m$ arises, taking the form 
\begin{equation}
    \vec{ J}_m=\rho \vec{ v}=D \vec{ \nabla} \rho
\end{equation}
where $\vec{ v}$ is the velocity in the fluid with local $\rho$ density. When this equation is obeyed, we will refer to \emph{Fick regime}. More interestingly, in fluid systems, it is found that a heat current arises if a density gradient is present, as well as a mass current can take place if a temperature gradient exists. This is tantamount to state that heat and mass transport are not independent but coupled. In the first case, the effect is called \emph{Dufour effect}, in the second one, \emph{Soret effect}. The magnitude of these effects is usually very small and they are in fact very difficult to observe experimentally since other phenomena, like natural convection \cite{zimmermann2022predicting}, is usually strong enough to cancel their contributions. It is thus of great interest to determine a computational procedure to determine the coefficients describing these two effects. In this Section we will clarify the specific characteristics of the systems of interest, its phenomenology, and the fundamental constitutive assumptions underlying the following derivation.

Our object of study, consists in non-reactive \cite{commento} binary mixtures in the liquid phase. Therefore, considering a fixed volume $\Omega$, it is assumed that two chemical species, hereafter referred to as species 1 and species 2, are contained within it. Thus, for the $i$-th species, the number $N_i$ of particles with mass $m_i$ is fixed. Next, we define the number density of particles $n_i$ and the mass density $\rho_i=m_i n_i$ as follows:

\begin{equation}
n_i=\frac{N_i}{\Omega}\quad,\quad \rho_i=m_i n_i=\frac{M_i}{\Omega}
\label{defs}
\end{equation}

where $M_i=m_i N_i$ is the total mass of the $i$-th species contained in the volume $\Omega$, both densities remain constant over time.\par
Since the system is in the liquid phase, each particle has non-zero velocities $\vec{ v}_i$. Therefore, given a closed surface $\Sigma$ completely contained within the volume $\Omega$, we define the mass flux $\vec{ J}_{m,i}$ due to the $i$-th species through the surface in terms of the mass density $\rho_i'$ in the volume $V'$ enclosed by $\Sigma$. It is straightforward to show that a conservation law must hold 
\begin{equation}
\frac{\partial \rho'_i}{\partial t}+\vec { \nabla} \cdot \vec{ J}_{m,i}=0
\label{mass_cons}
\end{equation}
Similarly, because of the particle motion, a flux of kinetic energy (which is to say, a heat flux $\vec{ J}_{q,i}$) is strictly related with the mass flux: by defining the specific heat capacity of the $i$-th species as $\mathcal C_{v,i}$, a second conservation law is thus found
\begin{equation}
\frac{\partial \rho'_{q,i}}{\partial t}+\vec { \nabla} \cdot \vec{ J}_{q,i}=0
\label{heat_cons}
\end{equation}
where $\rho'_{i,q}=\mathcal C_{v,i}\rho_i T$ is the heat density related to the flux of the $i$-th species. \par
Eqs. \eqref{mass_cons},\eqref{heat_cons} represent very general conditions, holding in both equilibrium and non-equilibrium conditions. In the case of an equilibrium condition, they are trivially satisfied, since $\rho'_i,\rho'_{q,i}$ must be stationary and no fluxes are present. Among non-equilibrium conditions, two different regimes are to be considered: (a) a steady-state condition, where no time-dependence of the involved fields is present (therefore eqs. \eqref{mass_cons},\eqref{heat_cons} just dictate that both $\vec{ J}_{m,i}$ and $\vec{ J}_{q,i}$ are solenoidal) and (b) a non-steady state condition, where time-evolution of the fields must be instead considered. \par
Stressing the fact the in this study we will focus on case (b), it is apparent that in order to obtain a set of equations describing the dynamics of relaxation toward equilibrium, explicit expressions for the fluxes $\vec{ J}_{q,i},\vec{ J}_{m,i}$ in terms of observable, macroscopic variables must be adopted. To this aim, by non-equilibrium thermodynamics arguments developed in the linear response regime, Onsager demonstrated \cite{de2013non,kjelstrup2008non} that the heat and mass currents can be expressed as linear combinations of generalized forces given by $\vec { \nabla}(1/T)$ and $ (\vec { \nabla}\mu)/T$ which enter in a system of coupled equations
\begin{equation}
    \begin{cases}
    \displaystyle              
        \vec{ J}_{q,i}=-\frac{L_{qq,i}}{T^2}\vec { \nabla}T-\sum \limits_{j=1}^{2} \frac{L_{q\mu,ij}}{T}\vec { \nabla} \mu_{j} \\ 
        \displaystyle
        \vec{ J}_{m,i}=-\frac{L_{\mu q,i}}{T^2}\vec { \nabla}T-\sum \limits_{j=1}^{2}\frac{L_{\mu \mu,ij}}{T}\vec { \nabla} \mu_{j}\\
    \end{cases}
    \label{case_mixture_general}
\end{equation}
where $L_{qq},L_{q\mu},L_{\mu \mu}$ are the Onsager coefficients, while $\mu_i$ is the chemical potential associated with the $i$-th species. In general, the chemical potential $\mu_i$ is a function of the $\rho'_i$ and can thus vary in time as well as in space. However the exact form of this dependence crucially depends on the specific interaction among the components of the system, and must thus be assumed depending on the case. We will this issue below.\par
In the case of a binary mixture, the system in eq.\eqref{case_mixture_general} can be rewritten in a more compact form by noticing that, since the labelling of a species as 1 or 2 is arbitrary, the process of heat and mass diffusion can be studied in terms of just species, say $i=1$, diffusing in the (dynamic) environment represented by the other species. Moreover, if the concept of mass fraction \cite{de2013non}
\begin{equation}
    w_i=\frac{M_i}{\sum \limits_{i=1}^{2} M_{i}}
\end{equation} 
is introduced, the constraint $w_1=1-w_2$ must hold. Therefore, as also reported by Zimmermann \cite{zimmermann2022predicting}, the dynamics expressed in eq.\eqref{case_mixture_general} is completely summarized by focusing on the study of just $\vec{ J}_{q,1}$ and $\vec{ J}_{m,1}$. Neglecting the $i=1$ index, we thus have \cite{zimmermann2022predicting}
\begin{equation}
    \begin{cases}
    \displaystyle    
        \vec{ J}_{q}=-\frac{L_{qq}}{T^2}\vec { \nabla} T-L_{q\mu}\frac{\partial \mu}{\partial w_1}\frac{1}{(1-w_1)T}\vec { \nabla} w_1\\
        \displaystyle    
        \vec{ J}_{m}=-\frac{L_{\mu q}}{T^2}\vec { \nabla} T-L_{\mu\mu}\frac{\partial \mu}{\partial w_1}\frac{1}{(1-w_1)T}\vec { \nabla} w_1\\
    \end{cases}
    \label{case_mixture_binary}
\end{equation}
From a phenomenological point of view, an equivalent but more informative formulation of the eq.\eqref{case_mixture_binary} was derived by Trimble and Deutch \cite{trimble1971molecular} in the case of a liquid binary mixture, explicitly containing the thermodiffusion coefficient $D_T$, the diffusion coefficient $D$ and the thermal conductivity $\kappa$
\begin{equation}
    \begin{cases}
    \displaystyle    
       \vec{ J}_q=-\kappa \vec { \nabla} T-D_T T \vec { \nabla} \left(\frac{\mu}{T}\right) \\
    \displaystyle    
       \vec{ J}_m=-\frac{D_T}{T}\vec { \nabla} T-D T \vec { \nabla} \left(\frac{\mu}{T}\right)\\
           \end{cases}
    \label{currents}
\end{equation}
In this derivation, the chemical potential $\mu$ is the chemical potential of the binary mixture, defined as $\mu=\mu_1/m_1-\mu_2/m_2$. We stress here, since it will be important in what follows, that $\mu$ vanishes if $\rho_1+\rho_2$ is homogeneous \cite{trimble1971molecular}. The link existing between the experimental accessible quantities and the Onsager coefficient is expressed by
\begin{equation}
\begin{split}
    &\kappa=\frac{L_{qq}}{T^2}\\
    &D=\frac{L_{\mu \mu}}{\rho (1-w_1)T}\frac{\partial \mu}{\partial w_1}\\ & D_T=\frac{L_{\mu q}}{\rho w_1 (1-w_1)}\frac{\partial \mu}{\partial w_1}
    \end{split}
\end{equation}
The physical meaning of each coefficient is enlightened by examining some specific limiting situations. If no chemical forces are present, but a temperature gradient is applied, the system eq.\eqref{currents} predicts a thermal current corresponding to the Fourier regime and a mass current, deriving from the thermal motion, clarifying the origin of \emph{thermodiffusion coefficient} name for $D_T$. Instead, in the cases in which the temperature is homogeneous but a spatial variation of chemical potential is present, it is immediate to verify that a mass current as described by Fick law appears, depending on the diffusion coefficient: the motion of the particles leads, in this case, to a corresponding flux of heat.\par
It is thus clear that $D_T$ has a critical role in determining a coupling between mass and heat transfer: this is particularly evident in non-equilibrium steady state situations where a temperature gradient is fixed among the mixture but the mass fluxes are free to relax. Under this hypothesis, provided that mass density is homogeneous, at equilibrium $\vec{ J}_{m}=\vec{ 0}$ and from eq.\eqref{currents} it is found that the magnitude of chemical potential gradient and the temperature gradient are directly proportional
\begin{equation}
    \frac{|\vec { \nabla} \mu|}{|\vec { \nabla} T|}=\frac{D_T}{TD}=\frac{\mathcal{S}}{T}
\end{equation}
where $\mathcal{S}$ is called the Soret coefficient. Physically, the Soret coefficient determines the magnitude of the concentration gradient arising from the presence of the imposed temperature gradient (\emph{Soret effect}). It is also found that a reciprocal effect, i.e. the inset of a temperature gradient as response for an imposed concentration gradient can be observed: this effect, which is the necessary consequence for the presence of a coupling term, is called \emph{Dufour effect}.\par
\section{\label{sec:2}Derivation}
The starting point for the transient analysis of the temperature $T(\vec{ r}, t)$ and density field $\rho'(\vec{ r},t)$ in the case of a coupled mass and heat transfer is the formulation of thermal current $\vec{ J}_q$ and diffusion current $\vec{ J}_m$ as derived by Trimble and Deutch \cite{trimble1971molecular}, reported in eq. \eqref{currents}. As stated in Section II, we will focus only on the dynamic of one of the species.\par 
In principle, the evaluation of the parameters involved in eq.\eqref{currents} is feasible if four quantities are known for every point $\vec{ r}$ and time $t$: the two currents $\vec{ J}_m(\vec{ r},t)$ and $\vec{ J}_q(\vec{ r},t)$ as well as the scalar fields $T(\vec{ r},t)$ and $\mu(\vec{ r},t)$. This approach, however, is particularly difficult to implement, since the time evolution of currents is affected by thermodynamic noise due to the energy and density fluctuations of the system. In addition, it was pointed out in relatively recent papers that the definition itself of thermal energy current is tricky if manybody potential are used to simulate the interactions \cite{doi:10.1021/acs.jctc.9b00252}.\par
By combining the conservation laws stated in eq.\eqref{mass_cons} and eq.\eqref{heat_cons} with the form of currents in eq.\eqref{currents}, a system of two partial differential equations (PDE) is obtained
\begin{equation}
\begin{cases}
\displaystyle    
\frac{\partial T}{\partial t}=\alpha_T \nabla^2 T +\frac{D_T}{\mathcal C_v \rho}\left[\vec{ \nabla} T \cdot \vec{ \nabla} \left( \frac{\mu}{T}\right)+ T \nabla^2 \left(\frac{\mu}{T}\right)\right]\\
\displaystyle    
\frac{\partial \rho'}{\partial t}=\frac{D_T}{T}\nabla^2 T-\frac{D_T}{T^2}(\vec{\nabla} T)^2 +D\vec{\nabla} T \cdot \vec{\nabla} \left( \frac{\mu}{T}\right)+DT \nabla^2 \left( \frac{\mu}{T}\right)
\end{cases}
\label{combine}
\end{equation}
where $\alpha_T=\kappa/\mathcal {C}_v \rho'$ is the thermal diffusivity.\par
In order to obtain the evolution of the mass density field, an expression for the chemical potential must be adopted. Here we use the fact that the system we are treating is a Lennard-Jones liquid. Assuming the leading order of all the main possible models \cite{STEPHAN2020112772}, the  dependence of chemical potential $\mu$ from $\rho'$ can be stated in the form
\begin{equation}
\mu=\mu_{0}+RT\ln\left(\frac{\rho'}{\rho'_0}\right)
\label{chem}
\end{equation}
where $R$ is the gas constant while $\mu_0$ and $\rho'_0$ refer to the chemical potential and mass density evaluated at an arbitrary reference point ($\mu_0$ can be eventually also set to zero).\par
Under eq.\eqref{chem} assumption, eq.\eqref{combine} can be rewritten in terms of $\rho'$. In fact, since $\vec{\nabla}(\mu/T)=R\vec{\nabla}\rho'/\rho'$, we have for the evolution of temperature field
\begin{equation}
\frac{\partial T}{\partial t}=\alpha_T \nabla^2 T +\frac{R D_T}{\mathcal C_v \rho'^2}\left[\vec{ \nabla} T \cdot \vec{ \nabla} \rho'+ T \nabla^2 \rho'+T\frac{(\vec{\nabla} \rho')^2}{\rho'}\right]
\label{combine2a}
\end{equation}
while the evolution of concentration field is
\begin{equation}
\begin{split}
\frac{\partial \rho'}{\partial t}=\frac{D_T}{T}\nabla^2 T-\frac{D_T}{T^2}(\vec{ \nabla} T)^2 +\frac{DR}{\rho'}\vec{ \nabla} T \cdot \vec{\nabla} \rho'+\\ \frac{RDT}{\rho'} \left(\nabla^2 \rho' -\frac{(\vec{\nabla} \rho')^2}{\rho'}\right)
\end{split}
\label{combine2b}
\end{equation}
In the most general case, the system formed by eq.\eqref{combine2a} and eq.\eqref{combine2b} has not a closed-form solution. A general method to find this class of solutions is possible if the system is in the form
\begin{equation} 
    \begin{cases}
    \displaystyle    
        \frac{\partial T}{\partial t}=\alpha_T \nabla^2 T + \beta_1 \nabla^2 \rho'\\
        \displaystyle
        \frac{\partial \rho'}{\partial t}=\beta_2 \nabla^2 T + \alpha_D \nabla^2 \rho'
    \end{cases}
    \label{fine}
\end{equation}
where $\alpha_T$ is the thermal diffusivity, $\alpha_D=D$ is the mass diffusivity (or diffusion coefficient), while $\beta_1,\beta_2$ can be properly defined as coupling parameters responsible for Soret and Dufour effect, respectively. Indeed, the system formed by eq.\eqref{combine2a} and eq.\eqref{combine2b} can be recast in the form \eqref{fine} if only the linear terms in each expression are retained. This approximation is in line with the assumption of a linear response regime constituting the bedrock of the expression for the fluxes adopted.\par
The basic idea at this point, is to proceed using a standard analytic technique allowing to restate the eqs.\eqref{fine} in a decoupled form, where the presence of coupling is transferred in the coefficients appearing in the new decoupled system.\par
This is done by defining the $\vec{ w}$ vector $(T(\vec{ r},t), \rho'(\vec{ r},t))$: the system in eq.\eqref{fine} is eventually written as
\begin{equation}
\frac{\partial \vec{ w}}{\partial t}=\mathbb A \nabla^2 \vec{ w}
\end{equation}
where $\mathbb A$ matrix of coefficients can be diagonalized by performing the transformation $\mathbb D=S^{-1}\mathbb A S$. By indicating with $\lambda_1, \lambda_2$ the eigenvalues of $\mathbb A$ and with $ v$ the vector resulting from $S^{-1}  w$, eq.\eqref{fine} reduces to
\begin{equation}
\begin{cases}
\displaystyle
\frac{\partial v_1}{\partial t}=\lambda_1 \nabla^2 v_1\\
\displaystyle
\frac{\partial v_2}{\partial t}=\lambda_2 \nabla^2 v_2
\end{cases}
\end{equation}
where the eigenvalues are in the form
\begin{equation}
\lambda_{1,2}=\frac{(\alpha_T+\alpha_D)\pm \sqrt{(\alpha_T-\alpha_D)^2+4\beta_1\beta_2}}{2}
\end{equation}
Under the hypothesis of a uni-dimensional heat/mass transport (which, without loss of generality, we assume along the $x$ axis) in a rectangular simulation cell with length $L$ simulated in periodic boundary conditions in all the three dimensions, the separation of variable methods allows to found a solution for $v_{1,2}(\vec{ r},t)$ in the form
\begin{equation}
\begin{split}
    v_{1,2}(\vec{ r},t)=\sum \displaylimits_{m=1}^{\infty} \exp\left(-\frac{4m^2\pi^2\lambda_{1,2}}{L^2}t\right)\times \\\left[c_1(m) \sin\left(\frac{2\pi m}{L}x\right)+c_2(m) \cos\left(\frac{2\pi m}{L}x\right)\right]
\end{split}
\end{equation}
where $c_{1,2}(m)$ coefficients are fixed by the initial profile $\vec{ v}(\vec{ r},0)$. \par
A natural choice is to build the initial $v_{1,2}(\vec{ r} ,0)$ by imposing across the simulation cell step-like temperature and concentration profiles. Formally, we thus have
\begin{equation}
    \vec{ w}(\vec{ r},0)=\frac{1}{2}\begin{pmatrix} (T_1+T_2)+(T_1-T_2)\text{sign}(x) \\ 1+\text{sign}(x)\end{pmatrix}
\end{equation}
where the concentration of species 1 is normalized to 1.\par
The initial conditions set for $\vec{ w}$ are then imposed to $\vec{ v}$ by performing the linear transformation $\vec{ v}=S^{-1}\vec{ w}$. At this stage, coefficients $c_{1,2}(m)$ are then fixed by using the Fourier series expansion of $\text{sign}(x)$ 
\begin{equation}
\text{sign}(x) = \frac{4}{\pi}\sum \displaylimits_{m=0}^{\infty} \frac{1}{(2m+1)}\sin\left(2\pi\frac{2m+1}{L}x\right)
\end{equation}
allowing to define an analytical expression for the evolution of the temperature and density time evolution. In particular, the evolution of the temperature and concentration differences between the two regions [$-L/2,0$] and [$0,L/2$] of the simulation box are
\begin{equation}
\begin{split}
\langle \Delta T(t)\rangle=\frac{8\Delta T(0)}{\pi^2}\left [ \sum \displaylimits_{m=0}^{\infty}\frac{1}{a(m)^2} \exp\left(-\frac{4 \pi^2 a(m)^2}{L^2}\lambda_1 t\right)\right]\\
\langle \Delta \rho'(t)\rangle=\frac{8\rho'(0)}{\pi^2}\left [\sum \displaylimits_{m=0}^{\infty}\frac{1}{a(m)^2}  \exp\left(-\frac{4 \pi^2 a(m)^2}{L^2}\lambda_2 t\right)\right]
\end{split}
\label{new_AEMD}
\end{equation}

where $\Delta T(0), \rho'(0)$ denote the initial temperature and concentration difference, respectively, and $a(m)=2m+1$.\par
It is worth to note the strict analogy of the resulting equation for $T$ in eq.\eqref{new_AEMD} with the corresponding and well-established formulation derived in AEMD theoretical framework \cite{lampin2013thermal,melis2014calculating,CAPPAI2023}. In fact, if no coupling is present and mass diffusivity is negligible (as in solids \cite{Fugallo_2018,CAPPAI2023}) it is immediately to recognize that $\lambda_1=\alpha_T$, leading to a perfect correspondence to the new result we found with the usual AEMD formulation. Moreover, if no coupling is assumed and no thermal gradient is imposed, a striking analogy between temperature evolution and concentration evolution is found. We interpret this result as a consequence of the fact that Fourier and Fick law, under the constraints dictated by the conservation laws, determine the same PDE describing the evolution of a temperature and concentration field, respectively.\par
In this perspective, the eigenvalues $\lambda_{1,2}$ of the $\mathbb A$ transport coefficients matrix can be physically interpreted, if a coupling is present, in terms of \emph{generalized heat/mass diffusivities}. \par
A possible way to link this formal picture with the phenomenological Soret coefficent $\mathcal{S}$, describing the strength of thermophoresis phenomenon, is to recast the system in eq.\eqref{currents} as
\begin{equation}
\begin{cases}   
\displaystyle
    \vec{ \nabla} \left(\frac{1}{T}\right)=r_{qq}\vec{ J}_q+r_{qm}\vec{ J}_m\\
     \displaystyle
    -\frac{\vec{\nabla} \mu}{T}=r_{qm}\vec{ J}_q+r_{mm}\vec{ J}_m
\end{cases}
\end{equation}
where $r_{qq},r_{\mu \mu}, r_{q\mu}$ are the Onsager resistivities, which can be easily worked out from eq.\eqref{case_mixture_binary} algebraically.\par
In this framework, the Soret coefficient is written as the ratio between the thermodiffusion coefficient $D_T$ and mass diffusivity $D_m$ in the steady state regime
\begin{equation}
    \mathcal{S}=\frac{D_T}{D_m}=-\frac{r_{mq}}{r_{qq}}\frac{1}{c_m T (\partial \mu/\partial \rho')}
\end{equation}
Under the approximations previously stated on (i) the form of chemical potential and (ii) the form of the system \eqref{fine}, the approximate form is found
\begin{equation}
    |\mathcal{S}|=\frac{|\beta| }{\rho' T^2} 
\end{equation}
\section{\label{sec:3B}Computational setup}
\begin{figure*}
\includegraphics[scale=0.5]{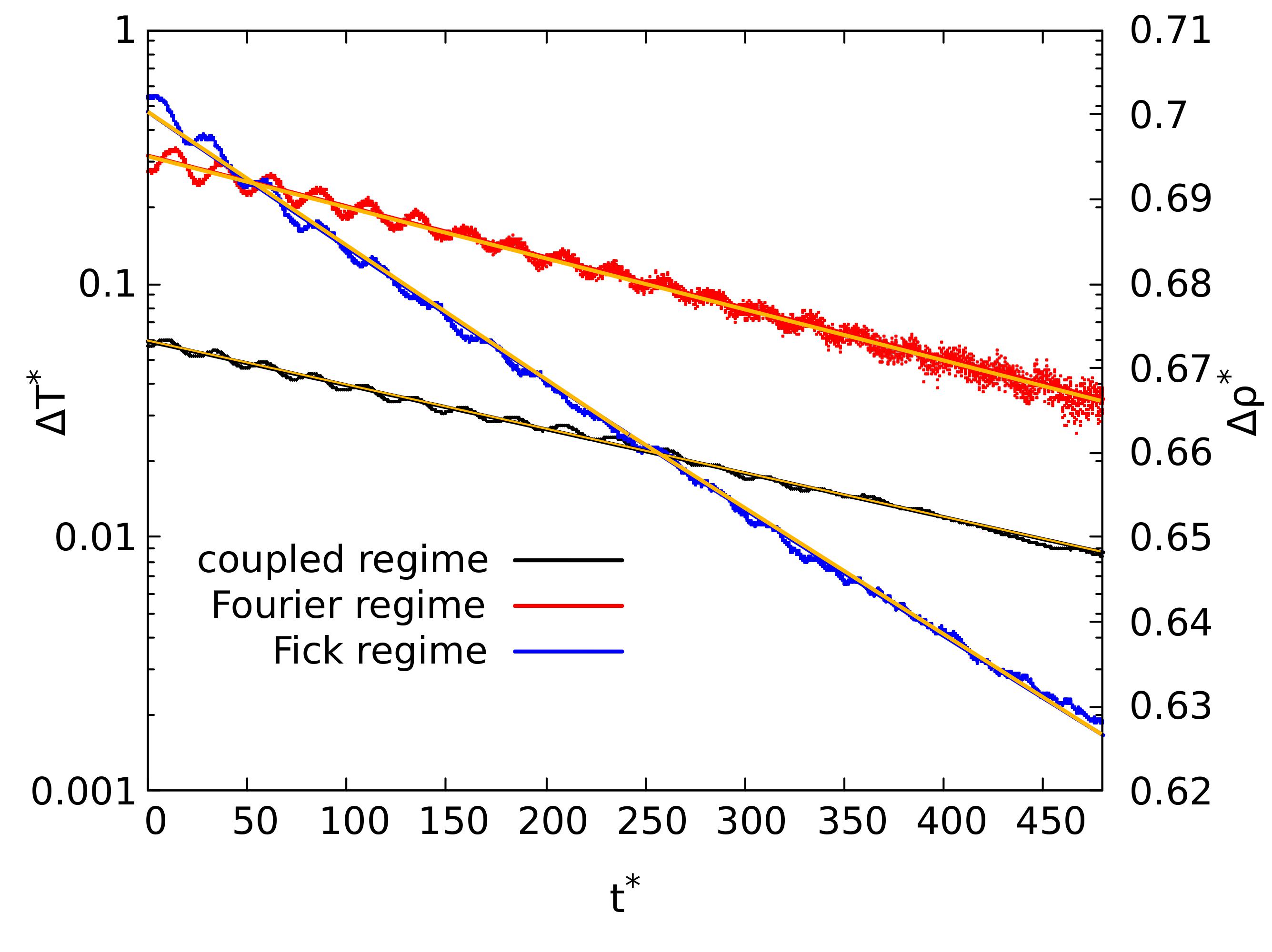}
\caption{Evolution of average temperature difference (left y-axis) and concentration difference (right y-axis) during the NVE run in the case of pure diffusion (Fick regime, blue), heat transport without concentration gradient (Fourier regime, red) and in the case of heat-mass coupled transport (black line). Instantaneous temperature and concentration differences monitored during the simulation are displayed as points, while the fitted trends are represented with solid yellow lines}
\label{fig:thermal}
\end{figure*}
Classical molecular dynamics (MD) simulations were performed using the LAMMPS software package. The system of interest is an equimolar Lennard-Jones liquid mixture composed of component 1 and 2, differing for the values of the LJ potential parameters ($\epsilon_1, \epsilon_2, \sigma_1, \sigma_2$) and the masses $m_1,m_2$. The values of the parameters and the masses are chosen in order to assure that both the pure phases as well as the mixture could simultaneously exists in liquid phase. Six different combination of mixtures are evaluated, as listed in Table \ref{tab:mixture1}.\par
The system contained a total of 86,000 atoms setting the initial density to $\rho^{*}=0.8$. An equal number of component 1 and 2 atoms were created in order to obtain a step-like concentration profile along the $L_x$ direction of the rectangular simulation cell. Periodic boundary conditions were applied in all directions but $L_x$. In order to prevent mass and heat diffusion before a complete equilibration of the system, two regions with width $2\sigma_1$ and encompassing the entire cross section of the simulation cell were created at the interfaces of component 1 and 2. Velocities and forces acting on the particle inside these two regions were kept to zero during the equilibration, which proceeded as follows.\par
The initial configuration is generated in its solid fcc phase and then minimized using the conjugate gradient (CG) algorithm until the pressure is 0. 
The bi-component system is then liquefied by a 10,0000-step NVE run at $T^{*}=5$. Subsequently, a 500,000-step NPT run is performed using Nosé-Hoover thermostat and barostat until average temperature and density are equilibrated at $T^{*}=0.8, \rho^{*}=0.8$, respectively.\par
The step-like temperature gradient is finally created by a 100,000 step NVT run. After this stage, which closes the equilibration run, all the atoms in the simulation cell are evolved in a NVE run. The number of steps was chosen to be sufficient to obtain a significant reduction of $\langle \rho_{1,2} \rangle$ and $\langle T \rangle$, being both these quantities evaluated during the run.
The thermal and mass diffusivities were calculated by fitting the concentration and temperature profiles obtained from the AEMD simulations using Eq.\eqref{new_AEMD}.\par 
In order to better estimate the $\beta$ values, for each mixture two additional separate runs were performed by setting (a) no initial thermal gradient (\emph{Fick regime}) and (b) no initial concentration gradient (\emph{Fourier regime}), the latter condition being realized by simply allowing a complete spatial randomization of the two components before monitoring.\par
In order to increase statistical significance of the obtained results, ten simulation for each mixture and for each transport regime were performed by varying the initial velocities and thermostatation run length. \par
\section{\label{sec:4}Results}

\begin{table}[t]
\centering
\begin{tabular}{ccccccc|ccc}
\hline
"DB"\\
\hline
Mixture &$\epsilon_1$ & $\epsilon_2$ & $\sigma_1$ & $\sigma_2$ & $m_1$ & $m_2$ & $\alpha_T$ & $\alpha_D$ &$|\beta|$ \\ 
\hline
1 & 1.0 & 0.6 & 1.0 & 1.1 & 1.0 & 0.89 &1.727 & 0.0659& $0.65\pm 0.9$  \\
2 & 1.0 & 1.1 & 1.0 & 0.9 & 1.0 & 1.1 & 1.459 & 0.0745 &$1.10\pm 0.8$  \\
3 & 1.0 & 0.8 & 1.0 & 1.1 & 1.0 & 1.33 & 1.755 & 0.0855 & $0.32\pm 0.60$  \\
4 & 1.0 & 0.9 & 1.0 & 1.1 & 1.0 & 1.33 & 2.898 & 0.0964 & $1.39\pm 0.72$  \\
5 & 1.0 & 0.9 & 1.0 & 1.5 & 1.0 & 1.33 & 2.889 & 0.543 & $0.38\pm 0.02$  \\
6 & 0.8 & 0.6 & 1.1 & 1.0 & 1.33 & 0.89 & 1.653 & 0.166 & $0.54 \pm 0.08$  \\
\hline
"RB" &  &  &  &  &  &  &  &  &  \\
\hline
Mixture &$\epsilon_1$ & $\epsilon_2$ & $\sigma_1$ & $\sigma_2$ & $m_1$ & $m_2$ & $\alpha_T$ & $\alpha_D$ &$|\beta|$ \\ 
\hline
1 & 1.0 & 0.6 & 1.0 & 1.1 & 1.0 & 0.89 &1.71 & 0.03& $0.60\pm 0.9$  \\
2 & 1.0 & 1.1 & 1.0 & 0.9 & 1.0 & 1.1 & 1.461 & 0.09 &$1.12\pm 0.9$  \\
3 & 1.0 & 0.8 & 1.0 & 1.1 & 1.0 & 1.33 & 1.721 & 0.1 & $0.27\pm 0.75$  \\
4 & 1.0 & 0.9 & 1.0 & 1.1 & 1.0 & 1.33 & 2.91 & 0.10 & $1.35\pm 0.81$  \\
5 & 1.0 & 0.9 & 1.0 & 1.5 & 1.0 & 1.33 & 2.89 & 0.57 & $0.31\pm 0.05$  \\
6 & 0.8 & 0.6 & 1.1 & 1.0 & 1.33 & 0.89 & 1.61 & 0.2 & $0.50 \pm 0.06$  \\
\hline
\end{tabular}
\caption{Simulation parameters for the six adopted equimolar liquid LJ mixtures, reported with the corresponding $\alpha_T$, $\alpha_D$ and $|\beta|$ values obtained using AEMD. "DB" and "RB" refers to direct and reverse temperature bias, respectively.}
\label{tab:mixture1}
\end{table}
The monitored time evolution of the average temperature and concentration differences (in adimensional LJ units) are schematically depicted in Fig.\ref{fig:thermal} for the paradigmatic case of mixture 1. In the same Figure, the analytical solutions Eq.\eqref{new_AEMD} fitted on the simulation results are represented, revealing an excellent agreement between the observed results. Results from fit for all the six mixture investigated are listed in Table \ref{tab:mixture1}.\par
From analysis of Fig.\ref{fig:thermal} several interesting conclusions can be extracted. As starting point, we observe that exponential decay prescribed by Eq.\eqref{new_AEMD} is substantially obeyed, providing a strong computational support for the theoretical derivation discussed in detail in Sec.\ref{sec:2}. The most evident deviation from the predicted exponential decay is represented by the presence of a clear sinusoidal overtone. This persistent feature is physically associated by a mechanical relaxation of the system strictly linked with the presence of the step-like temperature profile, determining a corresponding and unavoidable pressure gradient. In principle, this should require a re-setting of the theoretical background invoked to derive Eq.\ref{new_AEMD}, including the stress tensor contribution to the mass motion. In practice, we observe that the good agreement between the derived concentration and temperature trend can be read as an indication that in our investigated systems the mechanical contribution is sufficiently small to be neglected. This observation is also supported by two circumstances observed during the NVE run: (i) no significant increase of the average temperature of the system, indicating that the energy dissipated by the motion is negligible compared to the initial thermal energy of the system, (ii) direct inspection of the time evolution of average concentration differences reveal, in all the cases of study, a clear exponential decay.\par 
The results obtained for mixture 1, using the same parametrization of Zimmermann et al. \cite{zimmermann2022predicting}, are in accordance with their calculations: in particular, our values of thermal and mass diffusivity are respectively of $0.56\pm 0.11$ and $0.11\pm 0.2$, in good agreement with Zimmermann results ($0.48\pm 0.04$ and $0.087\pm 0.002$).\par
By repeating the set of calculation using the same initial concentration profile but with initial $T_1, T_2$ reversed (i.e. setting as cold the simulation half-cell that was set hot in the calculation for obtaining data in Table I), a set of diffusivity and coupling parameter was obtained, with magnitude perfectly compatible with the ones previously calculated.\par

\section{\label{sec:5}Conclusions}
In this paper, we reported a derivation of an approach to equilibrium method to estimate the coupling parameters of a binary LJ liquid. We have shown that, by combining the very general mass and energy conservation laws with the thermodynamics of non-equilibrium forms of heat and mass currents, a general system to formally decouple the transport phenomena can be found, if generalized diffusivity are introduced. Moreover, in the context of uni-dimensional flows and step-like initial profiles of temperature and mass density, a very simple solution for the evolution of the profiles can be found in terms of a sum of exponential. By applying this method to a LJ binary mixture, we provided a sound assessment of the method, comparing the results with the ones obtained with non-equilibrium techniques, and obtained as well the demonstration that no rectification effects seems to arise if initial temperature gradient is switched.
\bibliography{reference}{}

\begin{thebibliography}{10}
\bibitem{commento}
In~other words to keep the situation at the minimum level of complication it is
  assumed that no chemical reactions occur in~the system.
\end{thebibliography}
\bibliographystyle{unsrt}
\end{document}